\documentclass[aps,prl,twocolumn,showpacs,superscriptaddress,floatfix,nofootinbib]{revtex4}
\usepackage{graphicx}
\usepackage{epsfig}
\usepackage[english]{babel}
\begin{document}

\title{Quantum Phase Transitions in Anti-ferromagnetic Planar Cubic Lattices}

\author{Cameron Wellard}
\affiliation{Centre for Quantum Computer Technology, School of
Physics, University of Melbourne, Victoria 3010, Australia.}

\author{Rom\'an Or\'us}
\affiliation{Dept. d'Estructura i Constituents de la Mat\` eria,
Univ. Barcelona, 08028, Barcelona, Spain.}
\date{\today}

\begin{abstract}

Motivated by its relation to an $\cal{NP}$-hard problem, we analyze the ground
state properties of anti-ferromagnetic Ising-spin networks embedded on planar cubic lattices,
 under the action of  homogeneous transverse and longitudinal magnetic fields. 
This model exhibits a quantum phase transition at critical values of the
magnetic field, which can be identified by the entanglement behavior, as well
as by a Majorization analysis. The scaling of the entanglement in the critical 
region is in agreement with the area law, indicating that even simple systems 
can support large amounts of quantum correlations. We study the scaling
behavior of low-lying energy gaps for a restricted set of geometries, and 
find that even in this simplified case, it is impossible to predict the 
asymptotic behavior, with the data allowing equally good fits to exponential 
and power law decays. We can therefore, draw no conclusion as to the
algorithmic complexity of a quantum adiabatic ground-state search for the system.

\end{abstract}


\maketitle


Quantum many bodied systems are of increasing interest in modern
physics, particularly systems that exhibit a quantum phase transition
(QPT). The nature of the quantum correlations between the components
of these systems has been the subject of several recent studies 
\cite{Sa99, spin1, spin2, kike1, kike2, cirac, neill, qpt, qpt2, murg,
 nielsenqpt, jvidal, nordita, mac}, with the suggestion that systems in the vicinity of 
the critical point are highly entangled. 
In this work we consider the phase-structure of a spin-model for 
which finding the ground state can, 
for certain parameters, be proven to belong to the complexity 
class $\cal{NP}$-hard \cite{barahona}. 
We detect the presence of a QPT between paramagnetic and
anti-ferromagnetic phases and find that this transition 
is accompanied by a peak in the 
entanglement between different parts of the system with a scaling
behavior that makes it hard to simulate 
classically \cite{guifre, qpt, qpt2, guifre2}. Additionally, for the 
simplest realization of the spin network, we have calculated the minimum 
energy gap for networks of up to $N=24$ spins. The data accommodates 
equally good fits to both exponential and power law decays, allowing no 
conclusion to be formulated as to the efficiency of adiabatic quantum 
algorithms in solving this classically $\cal{NP}$-hard problem. Finally, we 
study the effects of frustration
in these spin networks, which adds an extra element of complexity, altering
the phase structure of the system.

\begin{figure}[h]
\centering
\includegraphics[width=0.3\textwidth]{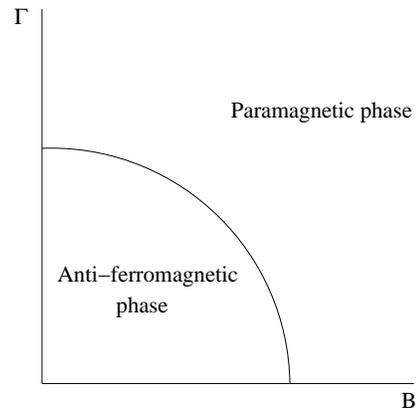}
\caption{Expected phase diagram in the ($B$,$\Gamma$) plane for the systems
  under study.}
\label{phased}
\end{figure}

The Hamiltonian under consideration is
\begin{equation}
H = \sum_{\langle i,j\rangle} \sigma_i^z \sigma_j^z + B \sum_{i = 1}^N 
\sigma_i^z + \Gamma \sum_{i = 1}^N \sigma_i^x \ ,
\label{ham}
\end{equation}
where the sums over $\langle i,j \rangle$ and $i$ run over the edges, 
and vertices respectively, of the particular
 lattice under study. The parameters $B$ and $\Gamma$ are respectively
the longitudinal and transverse magnetic fields, and $N$ is the number of
spins (qubits), each associated with an individual vertex of the lattice. 
 Spin networks of various planar cubic geometries are considered \cite{barahona}, and 
a QPT between paramagnetic 
(large fields), and anti-ferromagnetic (low fields) phases is expected 
in each case. The phase diagram of the system is shown schematically in  Fig.\ref{phased}.

A ground-state search for this system, over the entire set of planar 
cubic lattices, can be shown to be $\cal{NP}$-hard for the parameters $B=1,\Gamma=0$, due to a direct 
mapping to the problem of identifying the maximum cardinality of a stable set, 
for planar cubic graphs \cite{barahona}. In the presence of extremely high
fields however, the ground state is clearly trivial regardless of the size of 
the system, being simply given by the state in 
which all spins are aligned with the field. This gives rise to the possibility
of performing the, classically $\cal{NP}$-hard, ground state search quantum 
mechanically, using a quantum adiabatic algorithm. 
The procedure would be as follows: set the quantum register 
into the paramagnetic ground state of the system, with a 
very large magnetic field, and adiabatically change this 
field to the values $B=1,\Gamma=0$. Provided the evolution is adiabatic, the system is 
guaranteed to evolve to the ground state of this final Hamiltonian with a high probability. 
Clearly the evolution involves crossing the phase boundary, 
from the paramagnetic to the anti-ferromagnetic phase, and the time required 
to perform the procedure is determined by the energy spectrum at the critical point. 
It is, therefore, the scaling behavior of the energy spectrum at the critical
point, that determines the computational cost of the quantum adiabatic
ground-state search. This provides a strong motivation for studying the nature
of the QPT in this system.

The set of all possible planar cubic graphs grows rapidly with the number of 
nodes on the graph, with the behavior of different instances varying
considerably. We therefore find it convenient to begin the  paper by focusing 
on a particular, simple, instance of the planar cubic graphs,  which can be 
easily scaled. We then proceed to consider other particular instances, before 
finally considering the average behavior of the entire set. Our approach is 
numerical, and so we are restricted to small instances, $N\leq 24$ for even 
the simplest cases, making it difficult to draw any conclusions about the 
asymptotic behavior of the system.

\begin{figure}[h]
\centering
\includegraphics[width=0.4\textwidth]{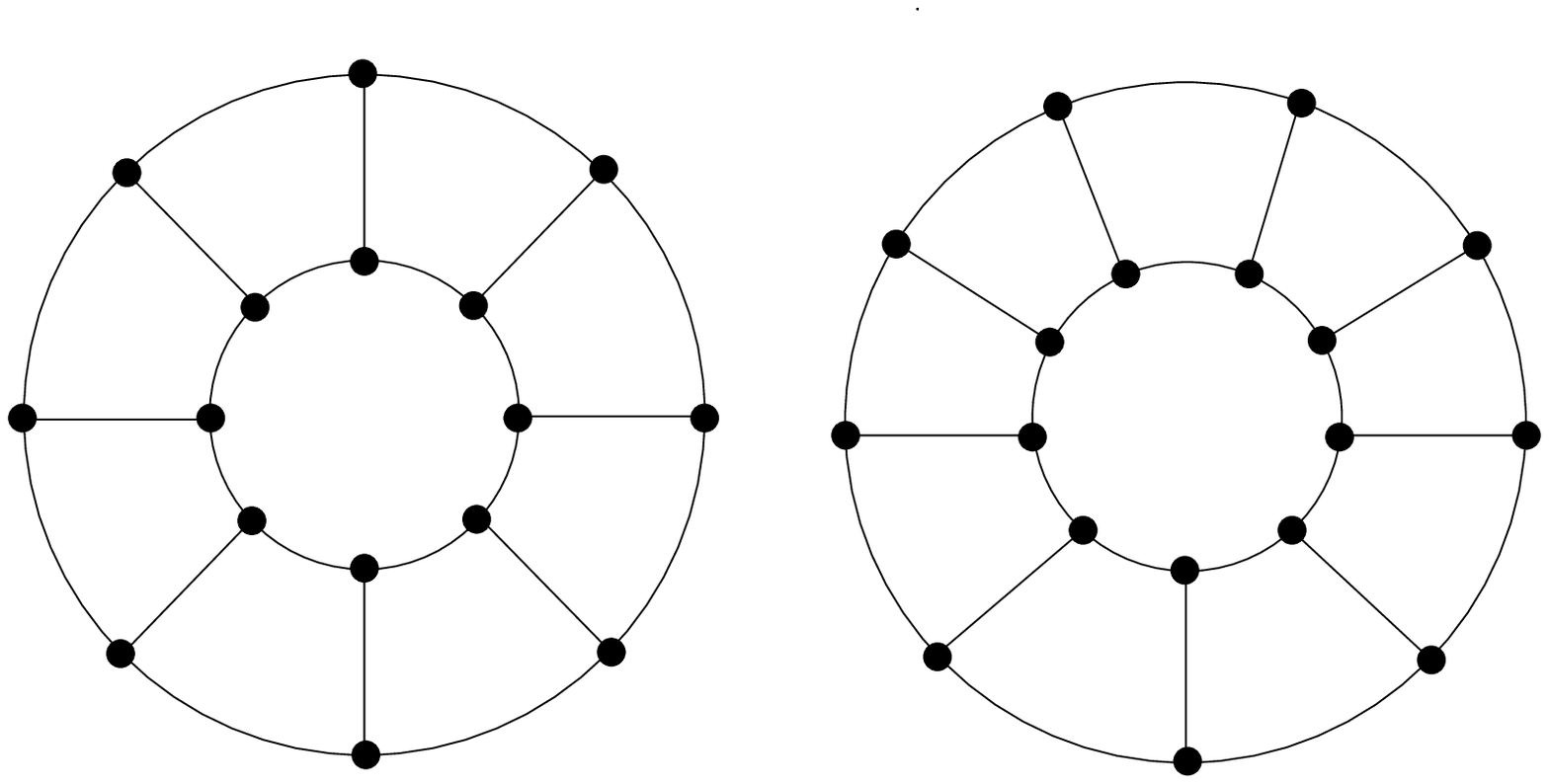}
\caption{``Ladder on a circle'' for $N = 16$ (left) and $N = 18$
  (right). For $N = 16$ the system is not frustrated, while for $N=18$
  it exhibits frustration.}
\label{ladder}
\end{figure}

Possibly the simplest subset of the planar cubic lattices that can be easily 
scaled, are those consisting of two coupled rings, shown in Fig.\ref{ladder}, 
which we refer to as the ``ladder on a circle'' geometry. As with cubic lattices in general, these
lattices always contain an even number of nodes $N$,  however, they can 
be divided into two classes. If there is an even number of nodes on each ring, 
the system contains no frustration, while if the rings contain an odd number
of nodes, then the lattice is frustrated. Although this frustration does not 
scale with the size of the system, the number of frustrated edges in the
ground state is two, regardless of the system size, and the two systems must 
have identical properties in the thermodynamic ($N \rightarrow \infty$) limit, 
for the small system sizes considered here, the frustration does have some effect.

\begin{figure}[h]
\centering
\includegraphics[angle=-90, width=0.5\textwidth]{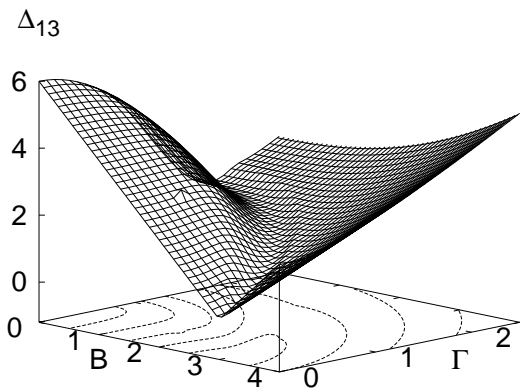}
\caption{Phase diagram of the non-frustrated ``ladder on a circle'', for $N
= 16$ spins. The energy gap between the ground and second excited
state reveals the appearance of two well separated phases. The critical line 
numerically fits the law $\Gamma = 1.93949 - 0.367302 B + 0.237182 B^2  - 0.106107 B^3$.}
\label{phasediagram}
\end{figure}

In the case of non-frustrated lattices the
phase structure of the system is evident even from the small lattice sizes 
considered in this article, $N\leq 24$, with clear indicators of a
QPT. Although the frustrated lattice must, in the thermodynamic limit, share 
the same phase diagram, the phase structure is more difficult to observe in small systems.
This point is illustrated by considering the low-lying energy gaps of the
system. Quantum phase transitions are marked by the vanishing of the first energy gap ($\Delta_{12}$), the difference
between the ground state and the first excited state energies. 
In the ``ladder'' geometry  this gap is finite in the paramagnetic phase, 
decreasing abruptly to zero, in the thermodynamic limit, for the
anti-ferromagnetic phase. In the non-frustrated system, the QPT is also 
marked by a clear minimum in the second energy gap $\Delta_{13}$, at the 
critical point, which can be easily observed for finite systems, as is 
shown in Fig.\ref{phasediagram} for $N=16$. For the frustrated ladder 
however, the multiple degeneracy of the ground state in the anti-ferromagnetic 
phase implies that the minimum will only be manifest in some 
higher energy gap, exactly how high is dependent of the size of the system, 
making it far more difficult to observe. 
 
Because of this compression of the energy levels at the critical points, 
it is the energy spectrum at this point that determines the minimum time-scale 
for adiabatic evolution between the two phases. In general the adiabatic 
time-scale is determined by the relation
\begin{equation}
\frac{|\langle e_n(t) | \partial H(t)/\partial t |e_1 (t)\rangle|}{\Delta_{1n}^2} << 1,
\label{eq:adiabatic}
\end{equation}
where $|e_n(t)\rangle$ is the $n^{\rm th}$ energy eigenstate of the
Hamiltonian $H(t)$. In the case of the non-frustrated ``ladder on a circle'', the lowest energy
level for which the matrix element in the numerator does not vanish, is the 
third energy level. In any case, because the ground state in the
anti-ferromagnetic phase is doubly degenerate, exciting the second energy 
eigenstate during adiabatic evolution from the paramagnetic phase does not
compromise the evolution. It is therefore the second energy gap, the energy 
difference between the ground state and the third energy eigenstate,  
which determines the minimum time-scale for adiabatic evolution in 
this system. The scaling of this energy gap with the size of the lattice, 
then, determines the computational complexity of a quantum adiabatic ground 
state search for these lattices. In the case of the frustrated ladders 
however, the multiple degeneracy of the anti-ferromagnetic ground state 
implies that the relevant energy gap is higher, exactly how high depends 
on the size of the system, and so it is far more difficult to determine 
the required adiabatic evolution time for such a geometry. 

We have analyzed the scaling behavior of the second energy gap 
for non-frustrated ladder geometries for system sizes up to $N=24$. 
The data allow equally good fits to either exponential or power law 
decays of the gap with lattice size, allowing us to draw no conclusion 
as to the asymptotic behavior of the scaling.  A confident determination
of the scaling law in this case would involve the analysis of much bigger
systems, which are clearly outside of our computational capabilities by 
the direct diagonalization techniques employed in this study. As a 
consequence, no conclusion can be formulated as to the efficiency of 
possible adiabatic quantum algorithms for a ground-state search for 
even this restricted subset of planar cubic lattices. The problem of 
computing the quantum adiabatic complexity of the the classically 
$\cal{NP}$-hard  ground state of generic anti-ferromagnetic planar cubic
systems with $B=1$ and $\Gamma = 0$ \cite{barahona, farhi1} is even 
more difficult. In this case the difficulty of determining a general 
scaling law for the energy gap using only a restricted number of 
small lattices is compounded by the fact that it is 
not {\it a priori} evident which energy gap determines the 
minimum evolution time. Different instances of planar cubic 
graphs, of the same size, can have vastly different degrees 
of frustration, and so different ground state degeneracies. 
The energy gap that determines the minimum adiabatic evolution 
time will therefore, in general, depend on the particular 
lattice under consideration.


The two magnetic phases can also be distinguished by the entanglement characteristics 
of their ground states. In the paramagnetic phase,
the ground state is not entangled, whereas in the magnetically ordered phase, 
the ground state is highly entangled. The critical point is, therefore, marked 
by some change in the entanglement behavior. Indeed it has been argued that one of 
the signatures of a QPT is entanglement on all length scales at the critical 
point \cite{spin1, spin2, kike1, kike2, cirac, neill, qpt, qpt2, murg,
 nielsenqpt, jvidal, nordita, mac}, in rough analogy to classical correlation 
functions in a thermodynamic phase transition.

 In this discussion we use an entanglement measure defined as the entropy of the reduced density matrix, 
which gives an indication of the entanglement between different blocks 
of particles. For instance, the single particle reduced density matrix
obtained by tracing out all but one particle, will have an entropy which gives a measure of the level of
entanglement between that particle and the rest of the
system. Similarly, if we trace out half of the particles, the entropy of the 
reduced density matrix gives a measure of the total entanglement between the block of particles remaining, 
and those that have been traced out.

The ground state entanglement between a single spin and the remaining system, calculated in this way
for a non-frustrated ladder, rises monotonically  as the strength of the field 
is decreased, moving from the paramagnetic to the ordered phase, due the 
$\mathcal{Z}_2$ symmetry of the system in the $\Gamma \rightarrow 0$, $B
\rightarrow 0$ limit (the ground state for 
the ``ladder geometry'' is close to a GHZ state of $N$ qubits for low 
values of $\Gamma$ and $B$). This degeneracy leads to some numerical noise in 
the data. As any linear combination of degenerate eigenstates is also 
an  eigenstate, this implies that some quantities, including entanglement 
measures, which differ between the different ground states, become ill defined at 
this point, and depend on which state is chosen. Additionally,
although actually non-degenerate for finite $\Gamma$, in larger
systems the first energy gap may be too small for the diagonalization 
algorithm to resolve, leading similar problems for small values of
$\Gamma$. This noise may be seen in some of the plots presented 
in this article, but we point out that wherever it is observed, the limiting
behavior of the system is clear.

\begin{figure}[h]
\centering
\includegraphics[angle=-90, width=0.5\textwidth]{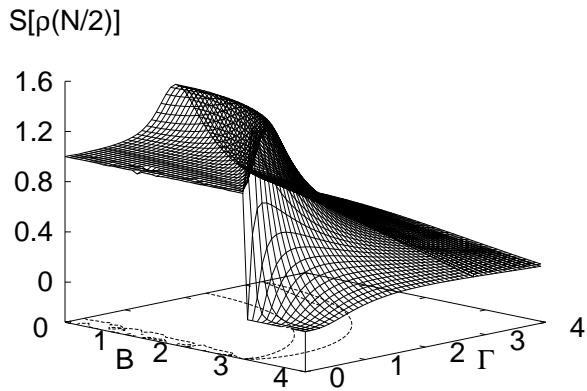}
\caption{Entanglement entropy of the reduced density matrix for the
particles on the inner (outer) ring, for $N = 16$. The peak indicates
the crossing of the critical region in the ($B$,$\Gamma$) plane. The critical 
line numerically fits the law $\Gamma = 1.93949 - 0.367302 B + 0.237182 B^2  
- 0.106107 B^3$.}
\label{entropy-N/2}
\end{figure}

In contrast to the single particle entropy, the QPT is indicated by a 
peak in the entropy of the reduced density matrix of $N/2$ connected particles 
$S(\rho_{N/2})$. In the case of the ladder geometries we have traced out the particles on one 
of the two rings, and the entanglement calculated in this fashion is shown 
in  Fig.\ref{entropy-N/2}. The QPT is therefore identified  by a collective measure of
entanglement, involving quantum correlations between large blocks of qubits.

\begin{figure}
\centering
\includegraphics[angle=-90, width=0.5\textwidth]{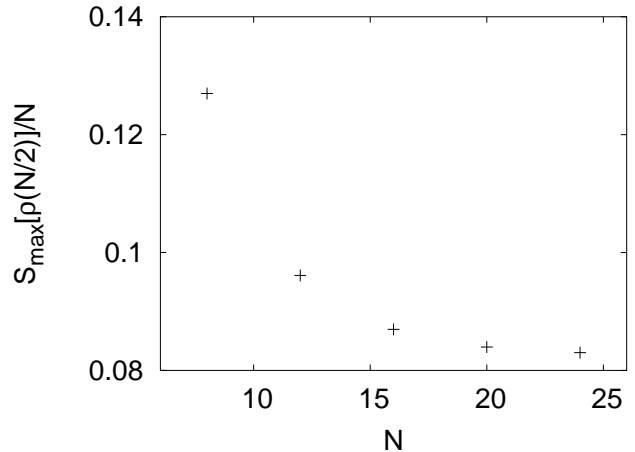}
\caption{Maximum entanglement entropy per site between two rings of a non-frustrated 
ladder configuration as a function of the number of lattice sites, up to
$N=24$, along the line $B=1$. For large $N$ the quantity $\frac{S_{{\rm
      max}}(\rho_{N/2})}{N}$ tends to a constant, which reveals an asymptotic
  linear scaling for the entropy peak.}
\label{scaling}
\end{figure}

The scaling of this entropy peak at the critical point with the size of the system, seems to be
in agreement with an area law which predicts that the entanglement between two blocks of 
spins should scale as the size of the boundary between the blocks, measured in terms of the
number of qubits. Note that, despite the fact that our configuration is
planar, the particular geometry and spin partition we have chosen forces the
scaling behavior of the entropy to be strongly linear, due to the 
linear number of short-range interactions we ``cut'' with the bipartition
between the inner and outer ring \cite{kike1, kike2, qpt, qpt2, cft2, cft3}. 
This result indeed implies that the ground state of the system at the critical 
point is close to a Valence Bond state, as
described in \cite{mac}.  
The scaling of this entropy peak for non-frustrated ladders, 
taken along the line $B=1$, indicates that the scaling tends 
to obey $S(\rho_{N/2}) \sim 0.08 \ N$ as the size of the system increases (see
Fig.\ref{scaling}). This behavior is similar to the linear law already found
in \cite{qpt} for the Exact Cover adiabatic quantum algorithm, and clearly
differs from the logarithmic law found in \cite{kike1, kike2} for quantum
spin chains.  Entanglement scaling according to
this particular bipartition of the system implies that even this
simple planar model is able to produce an exponentially large
amount of quantum correlations, as measured by the maximum rank of the
reduced density matrices over all possible bipartitions. This number can be
proven to be a measure of entanglement that controls the efficiency of
certain protocols in order to classically simulate the system in a conventional
computer. The results presented here imply that these classical simulation
protocols would become inefficient due to the large amount of entanglement present in the
system \cite{guifre, qpt, qpt2, guifre2}.

\begin{figure}
\centering
\includegraphics[angle=-90, width=0.5\textwidth]{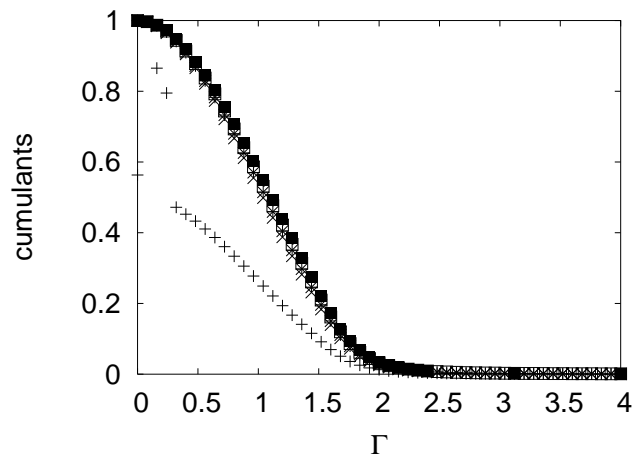}
\caption{Majorization arrow for $B=1$ when decreasing $\Gamma$, for $N
  = 16$. There is a change in the behavior of the cumulants at
  the critical point $\Gamma_c \sim 1.8$. For simplicity only the first five cumulants 
are plotted.}
\label{maj1}
\end{figure}

\begin{figure}
\centering
\includegraphics[angle=-90, width=0.5\textwidth]{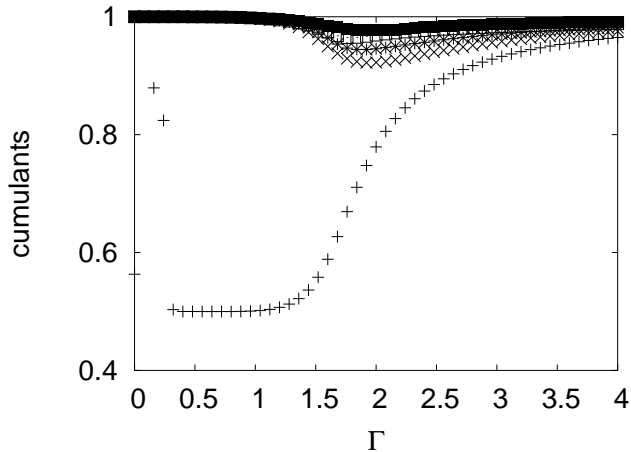}
\caption{Majorization analysis of the eigenvalues of the density
  matrix $\rho_{N/2}$ when tracing out one of the rings, when
  decreasing $\Gamma$, for $N = 16$ and $B = 1$. There is a 
  change in the behavior of the cumulants at
  the critical point $\Gamma_c \sim 1.8$. For simplicity only the first five cumulants are 
plotted.}
\label{majrho}
\end{figure}

Techniques from Majorization theory
\cite{muirhead, hardy, marshall, maj, nielsen-vidal}
have proven to be fruitful in characterizing the structure of the ground state
across a QPT. We have chosen to calculate the
$2^N$ cumulants $c_l = \sum_{i = 1}^l {\rm p}_i^{\downarrow}$
arising from the sorted probabilities ${\rm p}_i^{\downarrow}
 =  |\langle i |\psi_g\rangle|^2$ -where $|i\rangle$ is a quantum
state, expressed in the  $z$-basis (basis in which the Hamiltonian in Eq(\ref{ham}) is
written), 
$|\psi_g\rangle$ is the ground state of the
system, and ${\rm p}_i^{\downarrow}$ are the probabilities sorted into
decreasing order- at each step when varying the Hamiltonian
parameters. The Majorization analysis of this set of probabilities has
been useful in the study of efficiency in quantum
algorithms \cite{latorremaj, orus1, orus2}. Across the QPT on the line 
$B = 1$, we observe in Fig.\ref{maj1} that there exists a Majorization
arrow in the direction of decreasing 
transverse magnetic field $\Gamma$.
The phase transition is marked by a change in the cumulants at the critical
point, reflecting the transition from a paramagnetic to a magnetically ordered
phase. Step-by-step Majorization implies that the ground state is
becoming increasingly ordered after each step in a very strong sense,
at the precise level of each one of the probabilities, not just at
the global level of expectation values such as the mean magnetization.
An alternative Majorization analysis arises from the probability distribution
obtained from the eigenvalues of the reduced density matrix of $N/2$
particles, $\rho_{N/2}$ \cite{RG}. In the case of the ``ladder'' geometry we have
performed this analysis tracing out one of the two rings.
The cumulants calculated according to this procedure also change their behavior 
at the critical point, this time decreasing as the field is decreased 
across the QPT. This can be observed in Fig.\ref{majrho}.

\begin{figure}
\centering
\includegraphics[angle=0, width=0.5\textwidth]{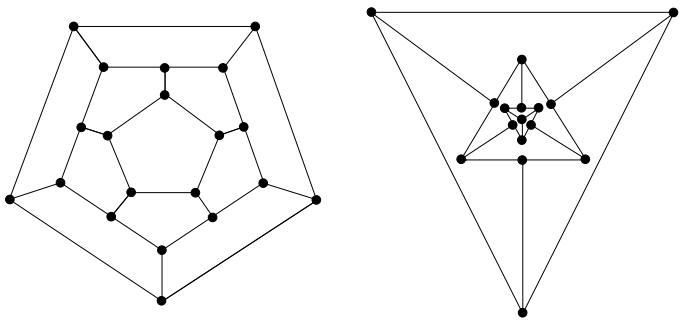}
\caption{Pentagonal (left) and tetragonal (right) geometries for
  cubic planar lattices of $N = 20$ and $N=16$ spins respectively.}
\label{penta}
\end{figure}

Other planar cubic lattices of different geometries
show similar behavior, with respect to these cumulants, in the critical region. For example, 
we have analyzed the two geometries
shown in Fig.\ref{penta} (which we call ``pentagonal'' and ``tetragonal''
geometries) for $N = 20$ and $N=16$ qubits respectively. These
 geometries are easily scalable,
and for small instances, highly frustrated, although  
the proportion of frustrated edges asymptotes to zero in the
thermodynamic limit.  
These lattices also exhibit a QPT between paramagnetic and anti-ferromagnetic phases, corresponding 
to a compression of the high energy levels of the system at the points at which  
correlations, as measured by the entanglement entropy, are maximum \cite{nielsenqpt}.

\begin{figure}
\centering
\includegraphics[angle=-90, width=0.5\textwidth]{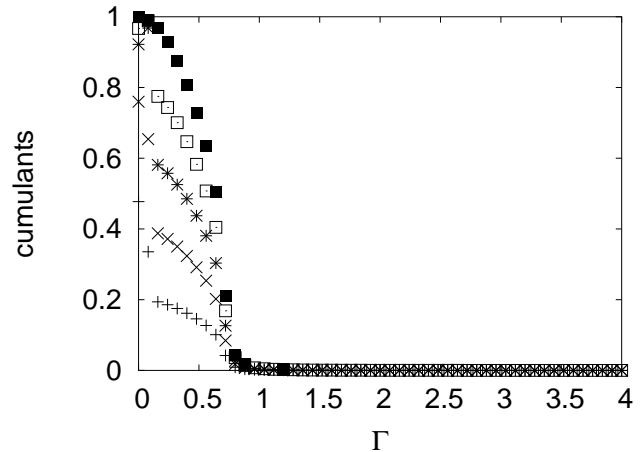}
\caption{Majorization cumulants for the pentagonal lattice of $N = 20$
  spins, for $B=1$ when decreasing $\Gamma$. A 
  change in their behavior is observed at the critical
  point $\Gamma_c \sim 0.7$. For simplicity only the first five cumulants are plotted.}
\label{majpent}
\end{figure}

\begin{figure}
\centering
\includegraphics[angle=-90, width=0.5\textwidth]{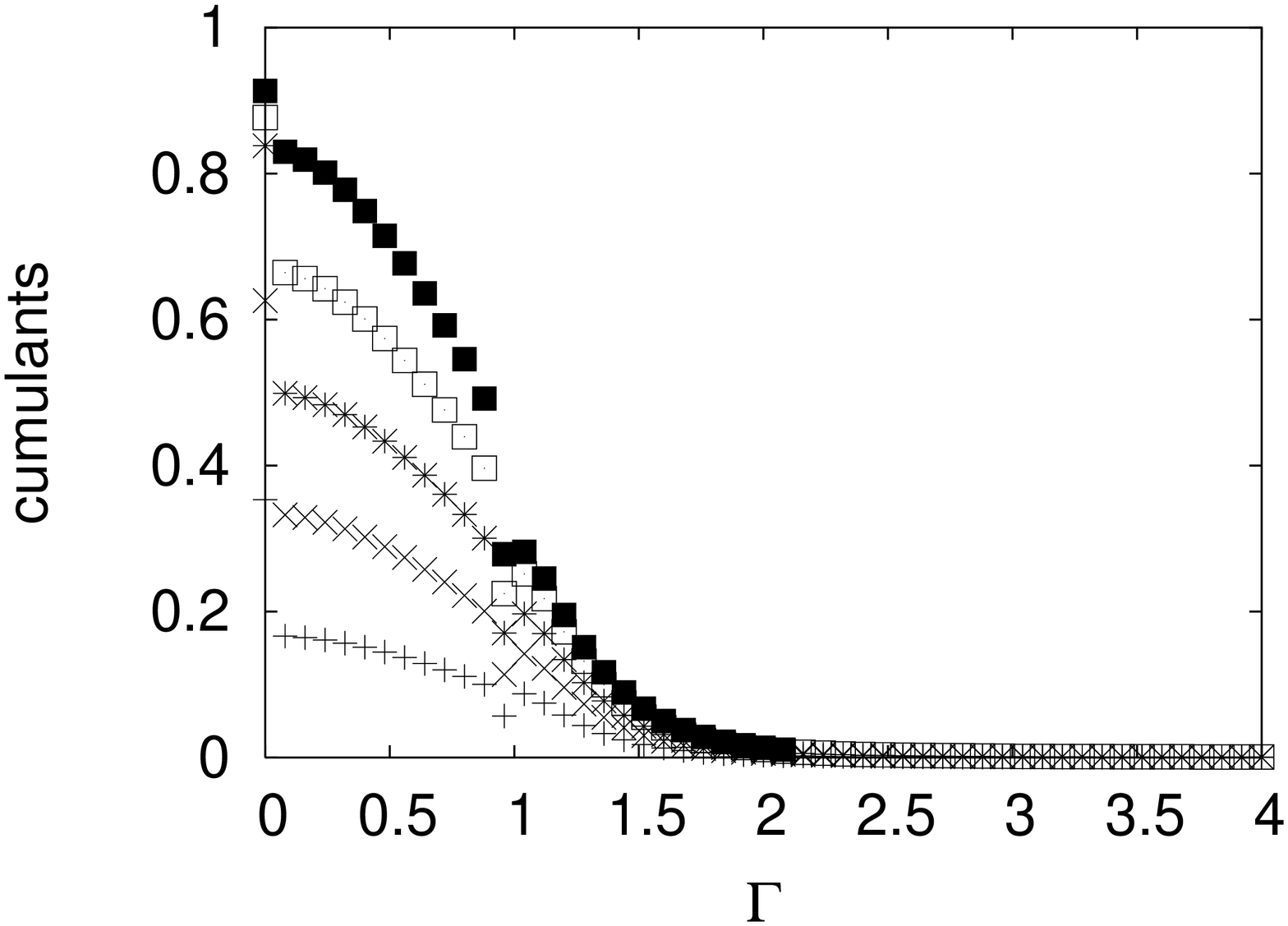}
\caption{Majorization cumulants for the tetragonal lattice of $N = 16$
  spins, for $B=1$ when decreasing $\Gamma$. A
  change in their behavior is observed at the critical
  point $\Gamma_c \sim 1.6$. For simplicity only the first five cumulants are plotted.}
\label{majtet}
\end{figure}

\begin{figure}
\centering
\includegraphics[angle=-90, width=0.5\textwidth]{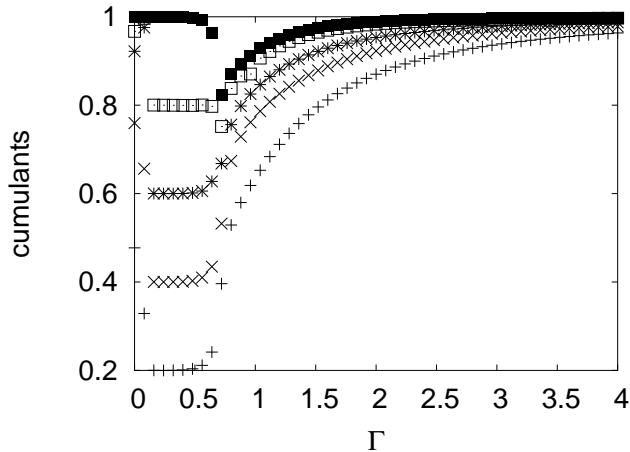}
\caption{Majorization analysis of the eigenvalues of the density
  matrix $\rho_{N/2}$ when tracing out half of the system, when
  decreasing $\Gamma$, for $N = 20$ and $B = 1$ on the pentagonal
  lattice. There is again a change in the behavior of the cumulants at
  the critical point $\Gamma_c \sim 0.7$. For simplicity only the first five cumulants are 
plotted.} \label{majrhopent}
\end{figure}

\begin{figure}
\centering
\includegraphics[angle=-90, width=0.5\textwidth]{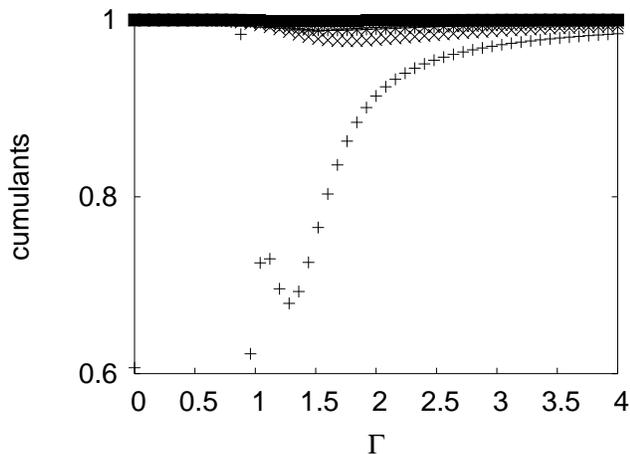}
\caption{Majorization analysis of the eigenvalues of the density
  matrix $\rho_{N/2}$ when tracing out half of the system, when
  decreasing $\Gamma$, for $N = 16$ and $B = 1$ on the tetragonal
  lattice.  There is a change in the behavior of the cumulants at
  the critical point $\Gamma_c \sim 1.6$. For simplicity only the first five cumulants are
  plotted. Unavoidable numerical noise is present for the first cumulant close
  to $\Gamma = 1$.} 
\label{majrhotet}
\end{figure}

The Majorization analysis for these lattices reveals similar results to
the ones presented for the ``ladder'' geometry: for $B=1$
and when decreasing $\Gamma$, the cumulants obtained from the
probabilities for the ground state, in the z-basis, increase at the critical point, as is
observed in Fig.\ref{majpent} and Fig.\ref{majtet}. In contrast to this, and
in agreement with the results obtained for the ladder geometries, the
Majorization behavior of the eigenvalues obtained from the reduced 
density operator $\rho_{N/2}$,  decrease at the critical point, as is seen in
Fig.\ref{majrhopent} and Fig.\ref{majrhotet}.

\begin{figure}
\centering
\includegraphics[angle=-90, width=0.5\textwidth]{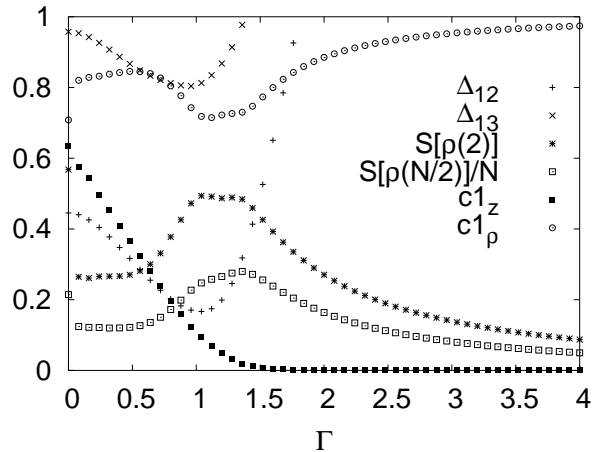}
\caption{Here we plot various parameters calculated for all 1249 possible 
planar cubic lattices with $N=18$, and averaged. The data were taken along 
the line $B=0.5$, and each parameter exhibits a signature of a quantum phase 
transition. The parameters plotted include the first and second energy 
gaps $\Delta_{12},\Delta_{13}$, each of which displays a clear minimum. 
The entropy of the reduced density matrix of a single spin 
(chosen at random) $S[\rho(1)]$ and a block of $N/2$ connected spins 
(again chosen at random)  divided by N $S[\rho(N/2)]/N$. Each of these 
quantities shows a peak at the critical region, between $\Gamma = 1$ and
$\Gamma = 1.4$. Finally we have plotted 
the lowest cumulants of a Majorization analysis based on the ordering 
of the ground state in the $z$ basis, $c1_z$, and based on an ordering 
of the eigenvalues of the $N/2$ particle reduced density 
matrix $c1_\rho$, The first of these shows a sudden increase, 
while the second exhibits a minimum on the vicinity of the QPT.}
\label{n=18_ave}
\end{figure}

To further illustrate our claims of Majorization signature for a QPT, 
we present results obtained using values calculated for all 1249 possible 
planar cubic graphs with $N=18$, and averaged. In Fig.\ref{n=18_ave} 
we show the lowest two energy gaps, the entropy of the reduced density
matrix of a single particle (chosen at random), the entropy of the reduced
density matrix of a block of $N/2$ connected spins (chosen at random), and the 
lowest cumulant as calculated using the Majorization procedures already
discussed. The data were taken along the line $B=0.5$, along which we expect a 
phase transition in all of the geometries. Our results show that all these
quantities change abruptly at the same value of $\Gamma$, strongly
suggesting the presence of a QPT.

\bigskip


To summarize, we have studied the quantum phase transition in anti-ferromagnetic planar cubic lattices 
under the
action of homogeneous magnetic fields. We have shown that techniques
from quantum information science \cite{nielsen-chuang}, such as the behavior of entanglement
entropy or Majorization theory, can be directly applied to the study
of quantum many-body systems, leading to new points of view in the
study of quantum critical phenomena. Lattices of different
geometries display
phase diagrams with characteristic and differentiated
anti-ferromagnetic and paramagnetic regions. Scaling laws for
entanglement at the critical point seem to
be in agreement with the area law, in a way that even the simplest
planar models are able to provide exponentially large quantum
correlations, as measured by the maximum rank of the reduced density
  matrices over all possible bipartitions. The understanding of the 
behavior of these systems is well
characterized by the use of Majorization theory and the analysis of
entanglement entropy, both of which reveal details about the
transitions present in the models which are sometimes
hard to be obtained by the study of the energy spectrum.

{\bf Acknowledgments:} R.O. acknowledges financial support from projects
MCYT FPA2001-3598, GC2001SGR-00065 and IST-1999-11053. The authors are 
grateful to the Les Houches School of
Theoretical Physics, where this work was initiated, and to discussions
with L.C.L. Hollenberg, T.D. Kieu, N. Lambert, J. I. Latorre, A. Prats and
E. Rico. Particular thanks to Wayne Haig 
(High Performance Computing System Support Group,
Department Of Defence, Australia) for computational support.


\end{document}